\documentclass{sig-alternate}
\begin{document}

\title{A Spin Glass Model for Human Logic Systems}
\numberofauthors{1}
\author{
\alignauthor F. Shafee\\
    \affaddr{Princeton University}\\
    \affaddr{Princeton}\\
    \affaddr{NJ, 08540, USA}\\
    \email{fshafee@princeton.edu}
     }
\maketitle
\begin{abstract}
In this paper, we discuss different models for human logic systems
and describe a game with nature.  G\"odel`s incompleteness theorem
is taken into account to construct a model of logical networks
based on axioms obtained by symmetry breaking.  These classical
logic networks are then coupled using rules that depend on whether
two networks contain axioms or anti-axioms.  The social lattice of
axiom based logic networks is then placed with the environment
network in a game including entropy as a cost factor. The
classical logical networks are then replaced with ``preference
axioms'' to explore the role of fuzzy logic.
\end{abstract}

\category{J.4}{ Social and behavioral sciences}{Economics}
\keywords{complexity, logic, entropy, spin glass}

\section{Introduction}

In this preliminary paper, we discuss the possibilities of
interactions between logic systems in complex agents placed in a
social network and between individual agents and nature. We
examine the effect of uncertainty in measurements by a neural
network.  First we consider a classical network with a fixed group
of axioms.  We discuss a quantum neural network that has no fixed
axioms, but the probability of using an axiom or its contradiction
with certain probability-coefficients in each ``decision making
measurement".  Then we investigate the interactions among these
logic networks in a social lattice, and later in a game with
nature.  We describe the role of entropy in the game with nature,
and in interactions with other agents.

\section{Review of Symmetry Breaking, Entropy and the Direction of Time}

Entropy was created in nature with symmetry breaking.  As the
ten-dimensional expanding universe stopped growing in certain
dimensions, momentons trapped in the stopped dimensions gave rise
to particles in order for their momentum modes to fit within those
defined dimension lengths. Entropy was thus created.  The second
law of thermodynamics states that entropy in a closed system must
always increase, attributing to the fact that nature would detest
particles clumping together by means of the gravitational force.
The gravitational force, on the other hand, is described as the
coupling between time and the spatial dimensions, and as the
universe expands, the gravitational coupling becomes weaker and
weaker, giving rise to increased entropy.

Hence time would become decoupled from space when entropy reaches its maximum.

\subsection{Information and Entropy}

Information may be described as a constraint in nature obtained by
symmetry breaking \cite{collier:entropy}.  An increase in
information would increase the complexity of the system or an
observer in the system because more pieces of information can give
rise to more and more laws in which the nature may evolve.

\subsection{Agents and Information}

When an agent obtains a piece of information from nature, it does so by breaking a symmetry.  This information gets
transformed into an axiom in the agent's cognitive network.  An increase in the amount of information and hence the
number of axioms will imply the possibility of more options by which the agent is able to make a decision by combining
the axioms with rules.

\subsection{Agents as Complex Entities}

An agent may be described as a complex meta-stable system that acquires information by interacting with nature.

We propose that agents are connected in social lattices.  It is possible to have clusters of social lattices that may
later be put into contact with one another.  An agent can process any information stored in its cognitive network.
Every agent may interact with nature macroscopically.  This interaction can be in the form of restructuring the agent's
environment.  Again, any reconstruction of the environment will imply an increase in entropy in nature.

\subsection{Definitions of Networks}

A classical logic network is a network that is based on a fixed set of axioms.  The axioms are connected with classical
gates or rules.  So given an input, the network uses the axioms and rules to produce an output or a decision.  All the
axioms and the gates in this model are fixed.  We can visualize the scenario by imagining a register representing an
axiom.  The register holds a value of either 1 or -1 depending on whether the register contains an axiom or the
contradiction of an axiom, which we shall refer to as an anti-axiom from now on.

A decision of the network consists of an ``action" which may be an
interaction with ``nature" at a macroscopic level, an interaction
with another logic network or an update of a coefficient with
which the network is connected to another network (this will be
discussed in detail later).

In the quantum version of the network, each register is replaced
with two registers that contain the coefficients for the axiom and
the anti-axiom, "a" and "b", such that $|a|^2+|b|^2$ = 1. However,
an agent's final decisions are based on a classical logical
semantics where classically defined axioms are connected by rules
to produce an output; so that we have a group ${(A=a_1,
a_2....a_n, R= r_1,r_2....r_n) \Rightarrow  D= decision}$.

During each decision making process, depending on the
coefficients, the (axiom + anti-axiom) function collapses to only
one of the states, taking the coefficients into account.  This
type of network would represent a model where a person has a
preference, but in different circumstances may choose differently,
perhaps based on the other inputs and using rules concerning
inputs and coefficients.  The improbability of quantum collapse in
the human brain has been explored \cite{tegmark:brain}.  So we
substitute the ``conscious" brain network with a model where we
introduce ``hidden variables."  In this model, a previous decision
or an environmental factor triggers the choice of an axiom over an
anti-axiom or vice versa by using the coefficient values stored in
the twin registers.  A detailed model of a "hidden variable"
driven "collapse" is being developed.

\subsection{The Classical Logic System and Incompleteness}

G\"odel's Incompleteness Theorem states that any classical logic
system must be incomplete \cite{godel:incompleteness}.  Classical
systems of logic are based on axioms, and the axioms themselves
cannot be proven within the logic system.

Now for any axiom in a logic system, we can define an axiom that contradicts it; namely, for C, we can define NOT(C).
Now a classical logical system may not have both C and NOT(C) as underlying axioms, as in that case the system will run
into inconsistencies when all the axioms are gated.  In other words, there may be no such condition as C AND (NOT(C)).
So the logical system chooses either C or NOT(C) as an axiom, and excludes its counterpart.

Now we try to define how quantum uncertainties and symmetry
breaking can give rise to logic systems that may be contradictory,
or in other words, logical systems that contain contradictory
axioms.  We propose that nature itself is initially apathetic to C
or NOT(C), or simply, nature does not prefer C or NOT(C).  So we
start from a logic vacuum, or a space where for every axiom there
is a coexisting anti-axiom.  In order to choose which of these
clauses should be picked as the underlying axiom, we resort to
symmetry breaking. We define states which are mixtures of
orthonormal axioms in the same manner we have mixed states in
quantum mechanics.

\subsection{Deriving Classical Logic Systems by Symmetry Breaking}

 It is possible to imagine a wave function of ``axioms" that may be ``collapsed" by some measurement.  The mode for this
 ``measurement" may be what Roger Penrose describes as ``consciousness" \cite{penrose:consciousness}.  More likely, it is
 an agent's interaction
 with nature that defines the rules, or etches the axioms into the neural network. So instead of a ``decision space",
 we actually go back to a wave particle duality space that establishes ``preferences" in the neural network.  In other
 words, we can say that an agent's perception organs acquire these axioms from nature by means of symmetry breaking.
 Again, each time a measurement is made, a symmetry is broken in the axiom space, so the axiom space is fixed inside a
 neural network.  However, this can be achieved only by disturbing the macro nature, or by increasing entropy in nature.
 So the axiom space becomes more defined by disturbing the macroscopic nature entropy.  We assume that this process of
 symmetry breaking and acquisition of information is local to the agent and its environment.

\section{Classical Logic Networks Placed in a Social Lattice}

Now we connect the logic networks in a model akin to the spin
model \cite{sherrington:spin}.  Here, we can draw similarities
with thermodynamics where a micro system (in this case the
perceptory organs - coupled to the neural network) is placed in
conjunction with a macro system (which is the environment).  These
logic networks placed in a society may be visualized as a small
thermodynamic spin lattice where the axioms residing in different
networks are coupled to one another by coupling constants
$J_{ij}$, and the lattice itself is coupled to a bigger lattice,
which is the environment.

However, since we can define the environment lattice to be huge compared to the neural network lattice, we can
probably take average values for interaction purposes and couple the lattice with the neural network lattice with
some multidimensional coupling factor.

In each of the lattices, spins are at a quantum level described as
``states" which can coexist in many orthonormal superpositions.
However, when the smaller lattice interacts with the bigger
lattice, the coupling causes the environment lattice to collapse
to a certain value. This value will depend on the probabilistic
coefficients of the wave functions and "most of the time" it would
yield the expected value. So an average person will end up with an
average set of axioms. Now each of the agents has a certain set of
axioms to start with.  Again, these agents are coupled with one
another in a lattice. A similar model with spin glass models and
evolution has been suggested at
$<http://pespmc1.vub.ac.be/SPINGL.html>$.  However, we argue that
the agents are not connected with one another with a random
coupling constant, $J_{ij}$, but some rules are defined, and also
that these $J_{ij}$s are updatable according to the specific state
of the entire network.
\\

1. $J_{ij}$ is not symmetric, i.e., $J_{ij}$ $\neq$ $J_{ji}$.  The
value of $J_{ij}$ depends on $i$ possessing axioms that
necessitate the existence of $j$. So $i$ will be coupled to $j$
more strongly if $i$ possesses axioms that require the existence
of $j$.

Now each agent will have the following behaviors in the game:
\\
1. Each agent $i$ will tend to flip its neighbor's axioms if the
neighbor's axioms contain contradictions of "i"'s axioms.  The
frequency and strength of flipping would depend on a coupling
constant $C_{ij}$.

$C_{ij}$ depends on the following:
\\
\\
a. The evolution of the logical code developed in $i$ that
contains that particular axiom,  i.e. the networking of the
certain axiom in the "logic network" of "i".  More accurately, the
number of decisions produced by $i$ that reflect the use of the
particular axiom.  In other words, $i$ will tend to flip a
neighbor's anti-axiom with more effort if the axiom has become an
important part of its network, and any future attempt of $j$'s
flipping it would cost $i$ dearly.

b. The determination of the number of the contradictory axiom in the neighboring agent's logic network.  An increased
frequency of anti-axioms in i's network would increase the possibility of an anti-axiom to be used in a future decision.

c. The effect of j's decision in i's environment (might be caused by physical distance between the two agents)

Now the total strength of coupling between i and j (i
$\rightarrow$ j) would be - $J_{ij}^n$ ($state_{ni}$ $\rightarrow$
$state_{nj}$)  +  $C_{ij}^m$ ($state_{mi}$ $\rightarrow$
$state_{mj}$), \\where $n$ and $m$ are states or registers
representing axioms.  We can implement this scheme by linking two
agents with appropriate gates.  The value of this coupling could
be described as "feelings" of agent $i$ towards agent $j$. This
total coupling will, at a macroscopic level, cause $i$ to play for
or against $j$, i.e. collaborate with $j$ or work against the
existence of $j$.

2. An axiom will flip if the effect of the neighbors' same axiom state coupling exceeds a flipping energy.  The flipping
energy depends on:

The certain axiom's connectivity with other axioms in the agent's cognitive network.

3. All agents must possess an axiom we shall call "self
preservation" or "preservation of the network" in random
probability.  We label this axiom $P$.  Agents containing neither
of these axioms cannot contribute to the existence of the agent or
to the network. In that case those agent's axioms, if few, will be
flipped by other agents; or they will self-destroy.  A later paper
will discuss the effect of agents possessing destructive axioms in
their cognitive network, or any critical number that will bound
the fraction of agents with self destructive axioms in a network.
However, in this paper, we assume that all agents possess $P$.

\section{Complexity versus Entropy}

An agent will try to increase its own or the network's complexity and hence stability by the acquired axioms as soon as
the acquired axioms get connected to the "existence axiom", E..  The agent's cognitive complexity increases as it
acquires more and more information from nature.  This complexity provides the agent with more and more options to
create a decision, and hence, increase the entropy of nature.

\section{Interactions between Agents and Nature: the Cost Factor}

Now each social lattice again is coupled to "nature", and we can assume the following game being played: The
"self preservation" clause makes an agent play against nature to preserve its own stability while nature tends
to increase disorder.  As disorder increases, so does entropy.  The agents play by acquiring information from
nature.  Axioms added to an agent's logical structure contribute to its network's complexity.  These new acquired
axioms again get entangled with the basic "self preservation" axiom.  This incidence adds more points to the
self-preservation side of the game, as now the agents have more rules they can use to preserve themselves.

Now we propose a game with the following rules:

Each agent has an axiom called "self preservation" or
"preservation of network" chosen at random. We call this axiom the
preservation axiom or $P$.

The other axioms are acquired by "interaction" with nature.

Acquiring each axiom has a cost factor $C$, since the process
requires measurements that increase entropy of the system.  The
increase of entropy disturbs the "meta-stable" agent state. The
acquired axioms get "gated" with the "preservation axiom" and add
points to further stabilize the agent. The acquired axioms in
different agents are again connected.  The contradictory agents
are connected repulsively and the supporting axioms attractively.

This game is now being simulated using different values for the
couplings constants and the costs. In an earlier work we have
presented simulations for a very simple network where agents are
connected to neighbors with unitary gates and are allowed to flip
when the effects from the neighbors cross a threshold
\cite{shafee:cnot}.

We add an extra dissipation term with each flip that is linear in
entropy, and also a coefficient term with each CNOT gate now
chosen at random.  In this very simple model, the spins form a
closed social network, and we are ignoring the acquisition of new
axioms from nature.

\section{THE Decisions}

An agent would use a subset of the axioms and rules to produce a decision.  The decision might be a macroscopic
interaction with the environment.  This may lead to restructuring the environment.  A restructuring event consists
of breaking an organized structure and creating a new one.  This action increases the entropy and decreases the total
free energy of the system as

\begin{displaymath}
F = H - TS.
\end{displaymath}

Now the interaction might reflect the presence of an axiom by
breaking a symmetry in the environment and might shift the
expectation value of the axiom from $<A>$ to $<A1>$.  This implies
that another agent seeking an axiom will now have a higher
probability of obtaining $A$ instead of $\bar{A}$ (Here, by $A$
and $\bar{A}$ we mean axiom and anti-axiom).  In such a case, an
agent possessing $\bar{A}$ will incur a cost factor because
\\
\\
1. Acquiring $\bar{A}$ has cost the agent. \\
2. $\bar{A}$ might have become connected with its preservation axiom.\\

The agent possessing $A$ has two options now:
\\
a. Inverting $\bar{A} $ to $A$\\
b. Trying to convert agents holding $A$\\

Whether an agent will choose strategy a) or b) will depend on the entropy cost.

\section{The Probability of Obtaining an Axiom by Another Agent}

The probability of obtaining an axiom or an anti-axiom by another
agent can now be written down in a simplified version by the
formula:\\

$F(A)=$ \\
$\sum C_{ijA} f_1(flip)- \sum J_{ijA}f_2(stabilize) -f_3(1/R)+ K
<A>$\\

Here, $A$ is the axiom, $R$ is a resistance factor for flipping
the axiom depending on how entangled the axiom is in the agent's
own cognitive network, $K$ is the coupling of the agent with
nature, and $<A>$ is the expectation probability of the axiom in
nature. We can see that this formula is very similar to the
formula for a classical neural network, except that the coupling
constants, unlike the weight factors in a regular neural network,
do not sum up to 1.  Also, instead of adding a term in $<A>$, it
might be more realistic to add a term $F(<A>)$. Here $F(A)$ is a
switching function that takes on a value of either -1 or 1,
depending on whether the RHS exceeds a certain threshold.  So
every time $A$ or $\bar{A}$ flips, the $C$'s and the $J$'s are
interchanged, and $R$ is updated to a new value that needs to be
updated with the accumulation of new axioms that get entangled
with the flipped clause.

A cost of entropy term must be added depending on whether the
agent has the "take entropy into account" axiom or the "ignore
entropy" axiom in its network.

\section{Trading Axioms}

Communicating axioms can be described as a method for obtaining these by several agents at a lower entropy cost than
would be necessary in an unsocial (single uncoupled spin) case.  The agent obtaining the axiom makes a decision and
interacts with its environment macroscopically so that it changes nature to have the information available at a lower
entropy cost by other agents.

This can be achieved only when a group of agents have a shared group of rules that relate symbols with possible axioms.

\section{Introducing "Fuzzy Axioms"}

A rigid axiom network in each agent will produce a system where too many conflicts are present.  We now define a system
where the axioms are fuzzy.  These "fuzzy" axioms may be described as preferences.

However, even in this system we initially keep the preservation
axiom, $P$, non fuzzy or stable.  The reason for this formulation
is as follows:  A "probabilistic" $P$ will cut the lifetime of the
agent, as every collapsed "lack of intention to exist" decision
will work against the existence of the agent.  Later in a
follow-up paper, we will present a simulation where we see the
effect of non-discretizing $P$.

So now we have a network with one register holding a value of 1
for $P$, and  qubits that represent the superposition of both the
axiom and the anti-axioms with coefficients a and be such that
$|a|^2+|b|^2=1$ for other axioms.

Now, in a logic system where we have one non-fuzzy axiom
coexisting with fuzzy axioms, the gates must be designed so that
no decision or input can modify the non-fuzzy axiom $P$.  Hence
all the other axioms may be connected in the logic network in a
way such that the acquired fuzzy axioms and the decisions or
interactions with other agents may modify the "preferences" or the
coefficients stored in the qubits, but $P$ stays connected to all
other axioms in a way such that no axiom can modify it.  The
connection path with $P$ must be one way.

Problems in this scheme arise when two agents are coupled, where
the coupling constants between the axiom lattices may tend to
modify $P$ for an agent as both agents try to independently
fulfill their axioms keeping only their own $P$ constant.

\subsection{Fuzzying the Existence Axiom}

In reality, an agent will interact with other agents, who will not
exist forever.  So the non-fuzzy "I exist" axiom will be coupled
with the information that agents die.  However, a flip in the "I
exist" axiom as an effect of adding this new piece of information
would destabilize the agent's cognitive network as any decisions
contrary to existence would be self-destructive.

\subsection{The Spurious Supporting Axioms}

An agent possessing $P$ in combination with the logical clause:
``I am an agent and agents die" will be more successful in
self-preservation if it can neutralize or alleviate the effect of
the latter clause by adding spurious axioms which would contribute
a positive factor to $P$ in order to stabilize it. The other
possibility may be "living in the present", i.e., ignoring the
cost factor of increased entropy and the future.

\subsection{Conflicting Spurious Axioms}

The spurious axioms that would support the ``I exist" axiom must
be non fuzzy, as fuzzy spurious axioms cannot always lead to the
same result, and may contribute negatively to the "I exist" axiom
at times.  So for the most efficient stabilization of $P$, the
designed spurious axioms must be non fuzzy. However, these
spurious axioms must be acquired in a fuzzy universe, and may
become also be coupled with fuzzy axioms.  So agents may again
possess a conflicting set of spurious axioms.

\subsection{The Spurious Axiom Game}

The spurious axioms game may be modelled as follows
\\
1. An agent will "believe" that the spurious axioms are true, and no contradiction may exist.\\

2. Agents containing a set of spurious axioms conflicting another agent's spurious axioms will try to flip the
second agent's spurious axioms.\\

3. If flipping is impossible, an agent may try to destroy another agent containing a conflicting spurious axiom as
the existence of any conflicting axiom may contradict the agent's self existence axiom by
coupling.\\

4. Another strategy would be to keep the spurious axioms private, without showing them to another
agent.\\

\subsection{Trading Spurious Axioms}

An agent unable to maintain a stock of spurious axioms that do not conflict with any other axioms in its cognitive
network may buy a spurious axiom from another agent.

The trade may consist of an axiom in exchange of a decision or
action by the buying agent.  The decision would cost work from the
agent, and hence the agent will lose time and increase the entropy
of its own local environment.  Agents selling spurious axioms will
seek to maximize their gain by selling the axioms to as many
agents as possible.

\subsection{Hiding Spurious Axioms}

An agent satisfied with its spurious axiom may hide it, as an open debate might render inconsistencies in it, and
hence oblige the agent to buy spurious axioms.

\section{The "Ignore Entropy" Axiom}

Any rule that would ignore the effect of future and entropy would ignore the cost factor in acquiring information,
and would make an agent maximize the acquisition of information from nature taking only the increasing complexity
factor of the agent into account, and not at the increasing entropy factor of the entire environment.  However, the
increased entropy of the entire environment again would affect all agents, including those who take the cost factor
into account.  Since "ignore the future" agents would destabilize the agents who take entropy into account, a conflict
would arise.

\section{The Game of "Ignore Entropy" and a Spurious Axiom}

An agent may play a game by superposing the "ignore entropy" axiom
with a spurious axiom.  This may reduce conflicts among spurious
axioms.  We try to formulate the possible games here and compare.
\\
Agent 1:  $S1$\\
Agent 2: $S2$ \\
Agent 3: $I$  \\
Agent 4: a $S1$ + b $I$   \\
Agent 5: c $S2$ + d $I$  \\
Agent 6: sell $S3$ but believe in $S4$ \\
Agent 7: confused; no axiom  \\

Here $S1$, $S2$ and $S3$ are spurious axiom 1, spurious axiom 2
and spurious axiom 3; $I$ is "ignore entropy"; and a, b, c and d
are coefficients such that in any one superposed wave, the sum of
the squares of the coefficients is 1.  Also, $S1$, $S2$ and $S3$
are normal to one another.  "$I$" introduces an extra cost factor
$K' (A)$ to other agents with every action made by an agent
holding $I$. Also, Agent 6 makes a profit of $K2$ by selling $S3$
to an agent. $K2$ is a function of the macroscopic work saved by
agent 6. However, Agent 6 must make an investment, $K4$, which is
a function of the work required to convert the agent.  This work
might consist of mapping $S1$ and $S2$ finding inconsistencies in
them.

Now Agent 1 and Agent 2 have two possibilities:
\\
a. Hide axiom\\
b. Show axiom\\

Besides holding $S1$ and $S2$, Agents 1 and 2 have two other
options:
\\
a. Sell axiom\\
b. Optimize on other axioms.\\

However, in order to "sell axiom", agents 1 and 2 must "show
axiom".  So "sell axiom" must be gated with show axiom with an
"AND".   Now if 1 and 2 optimize on other axioms, they can either
show axiom or sell axiom.  If they show axiom, 6 will now have its
cost function reduced, as 6 now only needs to find an anti axiom
to substitute into 1 and 2's axioms. If 1 and 2 hide axioms, then
6 will have to
\\
a) Find 1 and 2's axioms, and  \\
b) Find an anti-axiom.           \\

Now if 1 and 2 decide to show axiom and sell axiom then 1 and 2
will both try to flip each other and 6 will try to flip both 1 and
2. Given that 1 or 2 is able to flip the other and trade axioms, 1
or 2 will profit only when profit made from the other surpasses
the effort put into conversion. Now assuming that 1 and 2 both
have the same working ability, 1 or 2 will profit only when they
have to spend less than half the remaining lifetime of the other
trying to flip the other. Now calculating the cost factor and the
success of conversion can get very complex, and will be discussed
in a later paper.

Now agent 6 will have a strategy to find $S1$ and $S2$ properly
and design incoherence in them.  However, this strategy will
succeed depending on being successful achieving the following:
\\
a. Finding a subgroup of axioms in S3 which agents 1 or 2 may find more indispensable than $S1$ or $S2$ and which
conflicts with a subset of S1 or S2.\\
b. Convincing agents 1 and 2 that $S3$ has no logical flaws, based on the other axioms perceived by 1 and 2.\\
c. Not assimilating $S3$ into $S4$ and creating inconsistencies its own logical network.\\

Now any inconsistencies in $S3$ and the observation of $S4$ by
agent 1 or 2 will lead to disbelief in agent 6.  And since $S1$
and S2 are connected to existence axioms of agents 1 and 2,
inconsistencies between $S3$ and $S4$ observed by agents 1 or 2
will add a huge cost factor $K''$ to agent 6.

A more detailed simulation of the game is now on the way.

\section{Conclusion}

In this preliminary paper we merely suggest some modelling
possibilities for development of logical systems when human beings
interact. More detailed studies are being carried out.

\section{Acknowledgement}
The author would like to thank Andrew Tan for patient reading and feedbacks.

\bibliographystyle{abbrv}
\bibliography{axiom}
\end{document}